\documentclass{aa}
\usepackage{psfig,graphicx,ulem,epsfig}
\usepackage{txfonts}
\usepackage{float}

\begin{document}

\title{Link between brightest cluster galaxy properties and large scale
  extensions of 38 DAFT/FADA and CLASH clusters in the redshift range
$0.2<z<0.9$  }

\author{
F. Durret \inst{1} \and
Y.~Tarricq \inst{1,2} \and
I.~M\'arquez\inst{3} \and
H.~Ashkar \inst{4} \and
C.~Adami \inst{5}
}

\institute{
Sorbonne Universit\'e, CNRS, UMR~7095, Institut d'Astrophysique 
de Paris, 98bis Bd Arago, 75014, Paris, France
\and
Laboratoire d'Astrophysique de Bordeaux, Univ. Bordeaux, CNRS, B18N, all\'ee Geoffroy Saint-Hilaire, 33615, Pessac, France
\and
Instituto de Astrof\'isica de Andaluc\'ia, CSIC, Glorieta de la Astronom\'ia s/n, 18008, 
Granada, Spain
\and
Observatoire de Paris, UFE, 61 Avenue de l'Observatoire, 75014, Paris, France
\and
LAM, OAMP, P\^ole de l'Etoile Site Ch\^ateau-Gombert, 38 rue Fr\'ed\'eric 
Joliot--Curie,  13388 Marseille Cedex 13, France
  }

\date{Accepted . Received ; Draft printed: \today}

\authorrunning{Durret et al.}

\titlerunning{Brightest Cluster Galaxies}

\abstract
{In the context of large-scale structure formation, clusters of
  galaxies are located at the nodes of the cosmic web, and continue to
  accrete galaxies and groups along filaments. In some cases, they
  show a very large extension and a preferential direction.  Brightest
  cluster galaxies (BCGs) are believed to grow through the accretion
  of many small galaxies, and their structural properties are
  therefore expected to vary with redshift. In some cases BCGs show an
  orientation comparable to that of the cluster to which they belong.}
{ We analyse the morphological properties of 38 BCGs from the
  DAFT/FADA and CLASH surveys, and compare the position angles of
  their major axes to the direction of the cluster elongation at large
  scale (several Mpc).}
{The morphological properties of the BCGs were studied by applying the
  GALFIT software to HST images and fitting the light distribution
  with one or two S\'ersic laws, or with a Nuker plus a S\'ersic
  law. The cluster elongations at very large scale were estimated by
  computing density maps of red sequence galaxies. }
{The morphological analysis of the 38 BCGs shows that in 11 cases a
  single S\'ersic law is sufficient to account for the surface
  brightness, while for all the other clusters two S\'ersic laws are
  necessary. In five cases, a Nuker plus a S\'ersic law give a better
  fit. For the outer S\'ersic component, the effective radius
  increases with decreasing redshift, and the effective surface
  brightness decreases with effective radius, following the Kormendy
  law.  An agreement between the major axis of the BCG and the cluster
  elongation at large scale within $\pm 30$~deg is found for 12
  clusters out of the 21 for which the PAs of the BCG and of the
  large-scale structure can be defined. }
{The variation with redshift of the effective radius of the outer
  S\'ersic component agrees with the growing of BCGs by accretion of
  smaller galaxies from z=0.9 to 0.2, and it would be interesting to
  investigate this variation at higher redshift. The directions of the
  elongations of BCGs and of their host clusters and large scale
  structures surrounding them agree for 12 objects out of 21,
  implying that a larger sample is necessary to reach more definite
  conclusions.  }


\maketitle

\section{Introduction}

The brightest cluster galaxies (hereafter BCGs) are typically more than
one magnitude brighter than the second brightest galaxy and have long
been a topic of investigation, both based on observations and on
numerical simulations to explain their formation (see
e.g. Arag\'on-Salamanca et al., 1998). Their very high mass (typically
$10^{13}$~M$_\odot$) and extended stellar envelope suggest that they
have formed by the accretion of many galaxies (e.g. De Lucia \&
Blaizot 2007, Lavoie et al. 2016 and references therein).  The
alignments of the great axes of BCGs with the main directions of the
clusters to which they belong has been observed in a number of
occasions, suggesting that the accretion of smaller galaxies by the
BCG occurs predominantly along the direction of the cluster
elongation. This seems also to be the case for superclusters.  For
example, J\"oeveer et al. (1978) showed that in the Perseus--Pisces
supercluster the main galaxies of the clusters are directed along the
chain connecting this supercluster to other nearby superclusters. They
concluded that there must be a physical link between cluster galaxies
and their environment, suggesting a common origin and evolution of
galaxies and galaxy clusters in the cosmic web.  Many studies later
confirmed the presence of alignments of brightest cluster galaxies
(and sometimes of several bright cluster galaxies, not just the
brightest one) in all kinds of large scale structures (West \&
Blakeslee 2000, Plionis \& Basilakos 2002, Hopkins et al. 2005, Paz et
al. 2011, Tempel et al. 2015, Tempel \& Tamm 2015, Fo\"ex et al. 2017,
Hirv et al. 2017, West et al. 2017, Einasto et al. 2018).

Large-scale elongations and neighbouring clusters have been observed
around a number of clusters at various redshifts up to $z\leq 0.9$
based on density maps computed from catalogues of galaxies selected
around the cluster red sequence (Durret et al. 2016 and references
therein). In some cases, the extensions are much larger than the
typical sizes of clusters, suggesting we are detecting matter that may
be associated with the clusters and/or infalling into them, or in some
cases pairs of clusters or even superclusters. In view of the
alignments found between the BCGs and large scale structure mentioned
above, we propose to extend the sample of studied BCGs, by analysing
the properties of a sample of 38 BCGs covering the redshift range
$0.2\leq z \leq 0.9$, based on high quality HST images in the F814W
band.  Our aims are: 1)~to fit the BCGs with several models and see if
some of their properties vary with redshift, 2)~to compare the
position angles of their major axes with the general elongation of the
clusters derived from galaxy density maps of red sequence galaxies.

\section{Sample and data}

The sample of BCGs studied here includes 38 massive clusters. The
masses of the CLASH clusters are in the range 
$5\ 10^{14} < M_{vir} < 3\ 10^{15}$~M$_\odot$ (Postman et al. 2012)
and those of the DAFT/FADA have masses 
${\rm M_{200}}\geq 3\ 10^{14}$~M$_\odot$ (see e.g. Martinet et al. 2015). 

The BCGs cover the redshift range
$0.206\leq z \leq 0.890$ and have HST ACS imaging available in the
F814W band (except for ZwCl13332 which was observed in the F775W
band). Out of these, 21 clusters were taken from the DAFT/FADA survey
(Guennou et al. 2010)\footnote{\url{http://cesam.lam.fr/DAFT/index.php}} and
17 from the Cluster Lensing And Supernova survey with Hubble (CLASH)
survey (Postman et
al. 2012)\footnote{\url{https://archive.stsci.edu/prepds/clash/}}. Four
clusters are common to both surveys, in which case we used the CLASH
data. We limited our sample to the clusters with good quality data in
the BCG area, and eliminated those where a bright star or galaxy was
located too close to the BCG on the image.

For the DAFT/FADA clusters, the data were reduced with a modified
version of the HAGGLeS pipeline (Bradac et al. 2008) to subtract the
background and eliminate the bad pixels, and with Multidrizzle
(Koekemoer et al. 2011) to combine all the images for each cluster and
obtain a final image with a pixel size of 50~mas. 
For the CLASH clusters, the reduction was made by the CLASH team using
MosaicDrizzle (Koekemoer et al. 2011), leading to images with a pixel
size of 30~mas.  Since we are mostly interested here by the properties
of the BCGs at large scales, the fact that our images are sampled with
two different pixel sizes is not a problem.

\begin{table*}[h]
\centering
\caption{Sample of 38 clusters in which the BCG was studied in detail 
  and in a subsample of which the large scale structure around the cluster was 
  explored with a density map. The columns are : full cluster name,
  coordinates of the BCG, redshift, scale, relevant survey(s), and for the density maps: 
    instrument (M=CFHT/MegaCam,
    S=Subaru/SuprimeCam, V=ESO VLT/FORS2), filters. The coordinates 
  of BMW-HRI J122657.3+333253 are wrong in NED and have been corrected here.
  LCDCS 0829 is often also found under the name RX J1347.5-1145.}
\begin{tabular}{clllcccc}
\hline		
\hline		
Cluster           &  RA (J2000.0)  & DEC (J2000.0)& Redshift & scale  & Survey & Instrument & Filters \\
                  &                &              &          & (kpc/'') &      &            & \\
\hline		                                                     
Cl0016+1609       & $00:18:33.80$ & $+16:26:17.0$ & $0.5455$ & 6.385 & CLASH, DAFT/FADA & M & g',i'  \\
Abell 209	  & $01:31:52.55$ & $-13:36:40.5$ & $0.206$  & 3.377 & CLASH            & S & B,R \\ 
Cl J0152.7-1357   & $01:52:41.00$ & $-13:57:45.0$ & $0.8310$ & 7.603 & DAFT/FADA        & S & V,R \\
MACS J0329.6-0211 & $03:29:41.50$ & $-02:11:46.0$ & $0.450$  & 5.759 & CLASH            & S & B,I \\ 
MACS J0416.1-2403 & $04:16:09.13$ & $-24:04:03.5$ & $0.396$  & 5.340 & CLASH            & S & B,I \\
MACS J0429.6-0253 & $04:29:36.00$ & $-02:53:08.0$ & $0.399$  & 5.365 & CLASH            & S & V,I \\ 
MACS J0454.1-0300 & $04:54:11.10$ & $-03:00:54.0$ & $0.5377$ & 6.339 & DAFT/FADA        & M & g',z'\\ 
MACS J0647.7+7015 & $06:47:50.50$ & $+70:14:55.0$ & $0.584$  & 6.130 & CLASH, DAFT/FADA & S & V,I  \\ 
MACS J0717.5+3745 & $07:17:32.52$ & $+37:44:34.9$ & $0.548$  & 6.400 & CLASH            & S & V,I \\
MACS J0744.9+3927 & $07:44:52.80$ & $+39:27:26.6$ & $0.686$  & 7.087 & CLASH            & S & V,I \\
Abell 611     	  & $08:00:56.82$ & $+36:03:23.6$ & $0.288$  & 4.329 & CLASH            & S & B,R \\ 
Abell 851         & $09:42:58.00$ & $+46:59:12.0$ & $0.4069$ & 5.429 & DAFT/FADA        & M & g',i'\\
LCDCS 0110        & $10:37:52.36$ & $-12:44:49.0$ & $0.5789$ & 6.574 & DAFT/FADA        &   & \\
LCDCS 0130        & $10:40:40.27$ & $-11:56:04.2$ & $0.7043$ & 7.163 & DAFT/FADA        &   & \\
LCDCS 0172        & $10:54:24.42$ & $-11:46:19.4$ & $0.6972$ & 7.134 & DAFT/FADA        & V & V,I\\
LCDCS 0173        & $10:54:43.53$ & $-12:45:51.9$ & $0.7498$ & 7.337 & DAFT/FADA        &   & \\
CL J1103.7-1245   & $11:03:34.95$ & $-12:46:46.4$ & $0.6300$ & 6.835 & DAFT/FADA        &   & \\
MACS J1115.8+0129 & $11:15:51.91$ & $+01:29:55.0$ & $0.352$  & 4.958 & CLASH            & S & B,R \\
LCDCS 0340        & $11:38:10.18$ & $-11:33:38.1$ & $0.4798$ & 5.969 & DAFT/FADA        &   &\\
MACS J1149.6+2223 & $11:49:35.71$ & $+22:23:54.8$ & $0.544$  & 6.376 & CLASH            & S & B,R \\
MACS J1206.2-0847 & $12:06:12.15$ & $-08:48:03.4$ & $0.440$  & 5.685 & CLASH            & M & r',z' \\
LCDCS 0504        & $12:16:45.23$ & $-12:01:17.4$ & $0.7943$ & 7.490 & DAFT/FADA        &   & \\
BMW-HRI J122657.3+333253 & $12:26:58.25$ & $+33:32:48.6$ & $0.8900$ & 7.765 & CLASH, DAFT/FADA & S & V,I \\
LCDCS 0531        & $12:27:58.91$ & $-11:35:13.8$ & $0.6355$ & 6.861 & DAFT/FADA        &   & \\
LCDCS 0541        & $12:32:30.29$ & $-12:50:36.5$ & $0.5414$ & 6.361 & DAFT/FADA        &   & \\
MACS J1311-0310   & $13:11:01.80$ & $-03:10:39.7$ & $0.494$  & 6.065 & CLASH            & S & B,R \\
ZwCl 1332.8+5043  & $13:34:20.40$ & $+50:31:05.0$ & $0.6200$ & 6.786 & DAFT/FADA        & M & g',r' \\
LCDCS 0829        & $13:47:30.60$ & $-11:45:10.0$ & $0.4510$ & 5.767 & CLASH            & S & B,R \\ 
LCDCS 0853        & $13:54:09.75$ & $-12:31:01.4$ & $0.7627$ & 7.383 & DAFT/FADA        &   & \\
MACS J1621.4+3810 & $16:21:24.70$ & $+38:10:08.0$ & $0.4650$ & 5.867 & DAFT/FADA         & S & V,I \\
OC02 J1701+6412   & $17:01:23.00$ & $+64:14:09.0$ & $0.4530$ & 5.781 & DAFT/FADA       & M & g',i' \\
MACS J1720.2+3536 & $17:20:16.75$ & $+35:36:26.2$ & $0.391$  & 5.298 & CLASH            & S & B,I \\
ABELL 2261        & $17:22:27.10$ & $+32:08:02.0$ & $0.2240$ & 3.601 & CLASH            & S & B,R \\ 
MACS J2129.4-0741 & $21:29:26.20$ & $-07:41:26.0$ & $0.5889$ & 6.627 & CLASH, DAFT/FADA & S & V,I \\
RX J2129+0005	  & $21:29:39.96$ & $+00:05:21.2$ & $0.234$  & 3.722 & CLASH            & S & B,R \\ 
MS 2137.3-2353    & $21:40:15.16$ & $-23:39:40.1$ & $0.313$  & 4.586 & CLASH            & S & B,R \\
RX J2248-4431     & $22:48:43.97$ & $-44:31:51.3$ & $0.348$  & 4.921 & CLASH         &   & \\
RX J2328.8+1453   & $23:28:49.90$ & $+14:53:12.0$ & $0.4970$ & 6.084 & DAFT/FADA      & M & g',i' \\
\hline
\end{tabular}
\label{tab:sample}
\end{table*}

\begin{figure}
\begin{center}
\includegraphics[width=7cm, angle=0]{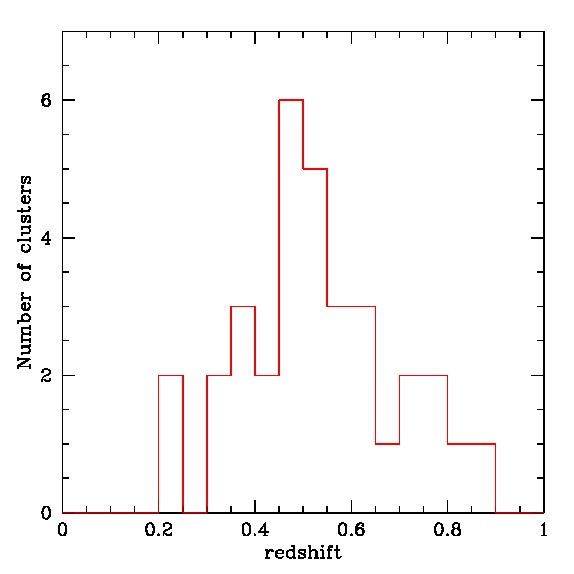}
\caption{Histogram of the redshifts for the 38 BCGs of the sample.}
\label{fig:histoz}
\end{center}
\end{figure}

The list of the 38 clusters analysed here is given in
Table~\ref{tab:sample}.  We give the full names of all the clusters in
column~1, but use abridged names in the text to make them more
readable. The coordinates given in Table~\ref{tab:sample} are those of
the BCG, which in a few cases differ slightly from those given in NED
for the cluster. The scales given in column~5 were computed with
  Ned Wright's cosmology
  calculator\footnote{\url{http://www.astro.ucla.edu/~wright/CosmoCalc.html}},
  taking H$_0=70$~km~s$^{-1}$~Mpc$^{-1}$, $\Omega_{\rm M}=0.3$ and
  $\Omega_{\rm vac}=0.7$. A histogram of the redshifts of the 38
BCGs of our sample is shown in Fig.~\ref{fig:histoz}.

\section{Surface brightness analysis of BCGs}

\subsection{The method}
\label{sec:method}

To fit the surface brightnesses of BCGs, we used the GALFIT software
developed by Peng et al. (2002).
Before running GALFIT, the knowledge of several parameters is
required. First, the PSF must be defined. This was done for all the
CLASH clusters by Martinet et al. (2017), who measured the FWHM of
stars on each of the images in the F814W band (using the PSFEx
software developed by
E.~Bertin\footnote{\url{http://www.astromatic.net/software/psfex}} and found
values between 0.05 and 0.25~arcsec (see Martinet et al. 2017,
Table~1). The properties of the HST ACS instrument in the F814W
  filter\footnote{\url{http://www.stsci.edu/hst/acs/documents/handbooks/current/c05\_imaging7.html}}
  leads to observe 80\% of the energy received in a radius of
  $\sim 0.2$~arcsec and 90\% in a radius of $\sim 0.5$~arcsec.  The
  values of the effective radius of the inner component $R_{e,int}$
  that we find (see below) are in the range 0.69 to 14.94~kpc,
  corresponding to 0.19 to 9.73~arcsec. Considering that the FWHM of
  the PSF is 2-3
  pixels\footnote{\url{http://www.stsci.edu/hst/acs/documents/isrs/isr0601.pdf},
    page 16}, for all the clusters with $R_{e,int}$ larger than 6
  pixels (twice the FWHM of the PSF) the influence of the PSF can be
  neglected, and there are only 6 clusters for which $R_{e,int}$ is
  smaller than 6 pixels.  We can note that this estimation is
  conservative, since some authors consider that the influence of the
  PSF can be neglected for clusters with $R_{e,int}>1$~pixel (see
  e.g. Hoyos et al. 2011). In this case, there is no need to take the
  PSF into account for any of our clusters.  We therefore decided not
to take the PSF into account in our analysis, since we mostly want to
study the shapes and extensions of the outer envelopes of the BCGs.

Second, the background of each image was measured in several zones
containing no visible object. Third, masks of bad pixels and of
neighbouring objects were constructed.

Several models were tested: a single S\'ersic, a double S\'ersic, a
Nuker plus a S\'ersic, and a core-S\'ersic model.

The S\'ersic law is defined as:
\begin{equation}
I(r)=I_{e} exp \lbrace -b_{n}\left[\left(\frac{r}{R_{e}}\right)^{1/n}-1\right]\rbrace
\end{equation}
\noindent
where $I_{e}$ is the intensity at the effective radius $R_{e}$ (the
half-light radius), $b_{n}$ is a constant defined by $b_{n}=2n-0.33$
(Caon et al. 1993) and $n$ is the S\'ersic index.

For a sample of BCGs in the redshift range $0.3<z<0.9$, Bai et
al. (2014) found that a single S\'ersic law with an index $n\sim 6$ was
sufficient to account for the profile. However, in some cases we could
not fit our BCGs with a single S\'ersic law, and we had to model the
BCGs by the sum of two S\'ersic laws, one accounting for the central
region, and the second one with the large envelope that characterizes
BCGs. The fits with two components were then of better quality.

Observations of BCGs with the HST have also indicated that these
galaxies often show a strong peak at their center. For example, for a
sample of 60 BCGs observed with the HST, Laine et al. (2003) found
that 88\% had well-resolved cores. In some of our BCGs, a central peak
was observed, so instead of a S\'ersic law we propose the Nuker model
to account for it, following the relation:
\begin{equation}
I(r)=I_{b}\ 2^{\frac{\beta-\gamma}{\alpha}}\left(\frac{r}{r_{b}}\right)^{-\gamma}\left[1+\left(\frac{r}{r_{b}}\right)^{\alpha}\right]^{\frac{\gamma-\beta}{\alpha}}
\end{equation}
\noindent
(Lauer et al. 1995), where $I_{b}$ represents the intensity at the
break radius that marks the transition between the inner and outer
power laws. The inner and outer power laws are described by $\gamma$
and $\beta$ respectively, and the $\alpha$ parameter allows us to
characterize the sharpness of the transition. Since the Nuker law
models the inner parts of elliptical galaxies, we added to a Nuker
profile a S\'ersic profile to account for the outer envelope.  We keep
in mind however the remarks of Graham et al. (2003) who noted that the
Nuker profile had five free parameters that could not all represent a
physical quantity.

We will therefore first start fitting the BCGs with a single S\'ersic
profile. If this fit is not good, we will try to improve it by adding
a second component, so that one S\'ersic law accounts for the outer
zone and another law for the inner zone. We will try both S\'ersic and
Nuker profiles for the inner component. Eleven cases can be fit with a
single S\'ersic component, the others require a two-component fitting,
and the Nuker inner profile provided results as good as a double
S\'ersic (see Tables 2 and 3) only in a handful of galaxies.

To decide if a fit is acceptable, we first look at its reduced
$\chi ^2$. As a second test to the quality of the fit, we create a
synthetic image of the galaxy built with the parameters of the best
fit and subtract it to the initial image to obtain an image of the
residuals. In some cases, we note that the residuals may be quite
large even if the reduced $\chi ^2$ is close to 1.0. As a third test
of the quality of the fit, we obtain the elliptical profile of each
BCG with the ELLIPSE routine in PyRAF, using the parameters determined
by GALFIT (such as the ellipticity and major axis position angle) as
guess parameters for the ELLIPSE routine. We then superimpose the
profile computed from the results given by GALFIT to the observed
profile. 

The residual and sharp divided images (see next subsection) are shown
and discussed in Appendix~A.  We do not show the profiles here to save
space.

\subsection{Sharp divided images}

A sharp-divided image (M\'arquez et al. 1999, 2003) is obtained by
dividing the original image, $I$, by a filtered version of it, $BI$,
in other words $I/BI$. In our case, the images are median filtered with the
IRAF command `median' using a box of 30 pixels. The result is very
similar to that of the unsharp masking technique (which subtracts $BI$
to $I$, that is computes $I-BI$, instead of dividing $I$ by $BI$), but
the former provides comparable levels for very different objects, so
the comparison between different objects is easier. Features departing
from axisymmetry, together with those with sizes close to the size of
the filter are better seen in the sharp-divided images. As seen in
Fig.~\ref{fig:cl0016rsd}, the results clearly
show several asymmetric structures in the center of a number of BCGs,
together with the presence of small companions close in projection,
that are not clearly seen in the original images. A few details on
individual BCGs are given in Appendix~A.

These sharp divided images can be compared to the residual images of
the 2D fit with two S\'ersic components.  When the model fittings do not
reproduce well the central regions and hence strong signal remains in
the residual images, sharp-divided images still show more details
since they are model-independent. Those cases generally indicate the
presence of additional components that the 2D fitting is not able to
reproduce (see for instance MACS0429, MACS0454, A851, Zw1332, RX1347,
MACS1621,OC02 or RX2328).

\subsection{Results}

\subsubsection{Modelling with a single S\'ersic function}

Eleven BCGs can be satisfactorily modelled with a single S\'ersic
function, and adding a second component does not improve the fit.  The
parameters of the fit are given in Table~\ref{tab:2sersic}.  For all
the other BCGs, it is necessary to add a second component. The
  parameters of the fits, exactly as provided in the output file from
  GALFIT, are given in Tables~\ref{tab:2sersic} and \ref{tab:nuk_ser},
  with the numbers of digits given by GALFIT. However, since the
  effective radii are computed in pixels, we prefer to give them in
  kpc and limit the precision on these quantities to one digit.  We
  note however that GALFIT substantially underestimates the true fit
uncertainties and its error bars are not very indicative of the true
error (see e.g. Haussler et al. 2007). This is the reason to avoid
considering them for plots; in fact, the errors on effective radii are
smaller than the symbols (see Fig. 2).

\subsubsection{Modelling with two S\'ersic functions}

\begin{table*}[h!]
\centering
\caption{Best fit parameters obtained for the sum of two S\'ersic models (38 BCGs): 
$n$ is the S\'ersic index, $m$ the magnitude (AB F814W) and $R_e$ the effective radius, with the 
$int$ and $ext$ indexes corresponding to the inner and outer profiles respectively. The last
column gives the reduced $\chi ^2$ of the fit.
For the clusters for which a single S\'ersic law is sufficient there is only one set of parameters.}
\begin{tabular}{cccccccc}
\hline		
\hline		
Cluster            & $n_{int}$ & $m_{int}$  & $R_{e,int}$ (kpc)& $n_{ext}$ & $m_{ext}$  & $R_{e,ext}$ (kpc) & $\chi^{2}_{\nu}$ \\ 
\hline		
Cl0016+1609        &  $3.26$  & $26.14$   &     $4.4$      &  $3030$  &  $25.10$  &   $45.0$  &  $0.245$  \\
Cl0152.7-1357      &          &           &                &  $3.07$  &  $21.62$  &   $3.2$   &  $0.511$  \\
Abell 209	   &  $0.81$  &  $20.09$  &  	$0.7$	   &  $2.41$  &  $15.60$  &   $29.2$  &  $1.625$  \\ 
MACS J0329.6-0211  &  $1.41$  &  $22.11$  &     $11.6$     &  $2.67$  &  $54.39$  &   $24.3$  &  $1.128$  \\
MACS J0416.1-2403  &  $1.24$  &  $18.78$  &  	$6.7$	   &  $1.92$  &  $16.84$  &   $118.3$ &  $1.856$  \\
MACS J0429.6-0253  &  $0.84$  &  $21.83$  &      $2.9$     &  $2.20$  &  $15.34$  &   $56.4$  &  $0.548$  \\
MACS J0454.1-0300  &          &           &                &  $4.40$  &  $24.93$  &   $42.2$  &  $0.262$  \\
MACS J0647.7+7015  &          &           &                &  $3.67$  &  $23.40$  &   $14.0$  &  $0.225$  \\
MACS J0717.5+3745  &  $0.52$  &  $22.19$  &  	$1.4$	   &  $2.42$  &  $18.30$  &   $21.0$  &  $0.979$  \\
MACS J0744.9+3927  &  $0.97$  &  $21.07$  &  	$4.4$	   &  $1.76$  &  $18.93$  &   $28.8$  &  $0.856$  \\
Abell 611     	   &  $0.82$  &  $19.77$  &  	$3.2$	   &  $1.88$  &  $16.14$  &   $35.5$  &  $1.038$  \\ 
Abell 851          &          &           &                &  $2.60$  &  $22.49$  &   $8.1$   &  $0.571$  \\
LCDCS 0110         &  $0.79$  &  $21.83$  &  	$1.3$	   &  $2.71$  &  $19.12$  &   $15.6$  &  $1.003$  \\
LCDCS 0130         &  $1.03$  &  $22.21$  &  	$2.1$	   &  $2.47$  &  $19.98$  &   $26.3$  &  $1.017$  \\
LCDCS 0172         &  $1.01$  &  $21.96$  &  	$1.3$	   &  $1.85$  &  $20.09$  &   $8.5$   &  $0.921$  \\
LCDCS 0173         &  $0.93$  &  $22.21$  &  	$2.6$	   &  $1.65$  &  $19.95$  &   $25.0$  &  $1.064$  \\
CLJ 1103.7-1245    &  $0.70$  &  $22.93$  &  	$1.3$	   &  $2.57$  &  $19.97$  &   $12.2$  &  $1.280$  \\
MACS J1115.8+0129  &  $0.90$  &  $20.25$  &  	$2.8$	   &  $2.47$  &  $16.79$  &   $60.9$  &  $1.060$  \\
LCDCS 0340         &  $1.42$  &  $20.82$  &  	$3.1$	   &  $1.55$  &  $19.47$  &   $16.8$  &  $0.977$  \\
MACS J1149.6+2223  &  $0.92$  &  $21.21$  &  	$4.2$	   &  $1.89$  &  $17.64$  &   $75.1$  &  $1.007$  \\
MACS J1206.2-0847  &  $2.41$  &  $18.47$  &  	$14.9$	   &  $2.46$  &  $17.36$  &   $158.8$ &  $1.105$  \\
LCDCS 0504         &  $0.77$  &  $22.00$  &  	$5.8$	   &  $1.40$  &  $19.59$  &   $27.8$  &  $0.937$  \\
BMW-HRI J122657.3+333253 &    &           &                &  $2.20$  &  $23.21$  &   $24.6$  &  $0.569$  \\
LCDCS 0531         &  $0.90$  &  $21.26$  &  	$2.4$	   &  $1.75$  &  $19.99$  &   $16.2$  &  $1.134$  \\
LCDCS 0541         &  $1.61$  &  $20.45$  &  	$5.0$	   &  $1.04$  &  $18.45$  &   $40.7$  &  $0.803$  \\
MACS J1311-0310    &  $0.89$  &  $20.66$  &  	$3.3$	   &  $2.38$  &  $17.91$  &   $34.4$  &  $0.978$  \\
ZwCl 1332.8+5043   &          &           &                &  $1.64$  &  $21.38$  &   $10.6$  &  $0.318$  \\
LCDCS 0829         &  $0.85$  &  $21.20$  &      $2.0$     &  $1.78$  &  $17.80$  &   $28.7$  &  $0.729$  \\
LCDCS 0853         &  $1.11$  &  $22.24$  &  	$2.5$	   &  $2.58$  &  $19.72$  &   $34.4$  &  $0.998$  \\
MACS J1621.4+3810  &          &           &                &  $2.00$  &  $19.67$  &   $4.0$   &  $0.544$  \\
OC02 J1701+6412    &          &           &                &  $2.63$  &  $22.49$  &   $14.9$  &  $0.528$  \\
MACS J1720.2+3536  &  $1.10$  &  $20.72$  &  	$2.0$	   &  $2.13$  &  $17.25$  &   $26.2$  &  $1.096$  \\
ABELL 2261         &  $0.51$  &  $17.70$  &  	$6.2$	   &  $1.26$  &  $15.34$  &   $27.8$  &  $5.398$  \\
MACS J2129.4-0741  &          &           &                &  $4.00$  &  $25.10$  &   $43.8$  &  $0.428$  \\
RX J2129+0005	   &  $1.10$  &  $19.00$  &  	$3.0$	   &  $1.92$  &  $15.58$  &   $50.5$  &  $1.088$  \\ 
MS 2137.3-2353 	   & 	      &  	  &  		   &  $2.31$  &  $16.70$  &   $18.2$  &  $1.128$  \\
RX J2248-4431      & $0.73$   &  $19.51$  &  	$4.6$	   &  $1.82$  &  $16.16$  &   $49.4$  &  $0.964$  \\
RX J2328.8+1453    &          &           &                &  $3.09$  &  $21.18$  &   $4.7$   &  $0.580$  \\
\hline		
\end{tabular}
\label{tab:2sersic}
\end{table*}

These last years two component models have been widely used to model
cluster cD galaxies. For example Gonzalez et al. (2005) used two de
Vaucouleurs laws, Seigar et al. (2007) and Donzelli et al. (2011)
considered the sum of a S\'ersic and an exponential law, while Madrid
\& Donzelli (2016) applied two S\'ersic laws.

This led us to fit with two laws the BCGs of our sample that could not
be fit with a single S\'ersic law.  The parameters of the fits are
given in Table~\ref{tab:2sersic}. For the inner component, we give the
index $n_{int}$, the corresponding apparent magnitude $m_{int}$, and
the effective radius $R_{e,int}$ in kpc.  The equivalent quantities
$n_{ext}$, $m_{ext}$, and $R_{e,ext}$ are given for the external
component, which corresponds to the extended envelope of the BCG. The
last column of the table gives the reduced chi-square
$\chi^{2}_{\nu}$, which indicates the quality of the fit.

There are several BCGs with $\chi^{2}_{\nu}$ values strongly differing
from 1.0, for which the fit cannot be considered as satisfactory,
based on the $\chi^{2}_{\nu}$ value alone.  For the BCGs of some other
clusters (e.g. MACS0717, LCDCS0172 and LCDCS0531), the
$\chi^{2}_{\nu}$ values close to 1.0 seem to indicate that the fit is
good, but the residuals and profiles do not appear acceptable.
This justifies the fact that we cannot base our decision on the
quality of fits only on the $\chi^{2}_{\nu}$ values. We will therefore
also base it on the residual images.

\subsubsection{Modelling with a Nuker and a S\'ersic functions}

\begin{table*}[h!]
\caption{Best fit parameters obtained for the sum of a Nuker and a S\'ersic 
models. The $\mu_{b}$ surface brightnesses correspond to AB magnitudes in the F814W filter. }
\begin{tabular}{cccccccccc}
\hline		
\hline		
Cluster &  $\alpha$ & $\beta$ & $\gamma$ & $\mu_{b}$ (mag/arcsec$^{2})$ & $R_{b,int}$ (kpc) & $n$ & $m$ & $R_{e}$ (kpc) & $\chi^{2}_{\nu}$ \\ 
\hline		
Abell 611     	   &  $2.16$  &  $1.89 $ & $0.06$ &  $19.92$          &  $1.9$           & $1.65$  &  $16.35$  & $37.3$ &  $1.035$ \\ 
MACS J1149.6+2223  &  $1.91$  &  $2.60$  & $0.01$ &  $21.36$          &  $3.3$           & $1.71$  &  $17.71$  & $78.4$ &  $1.005$ \\  
MACS J1206.2-0847  &  $1.29$  &  $2.37 $ & $0.01$ &  $20.13$          &  $1.3$           & $2.41$  &  $17.60$  & $40.9$ &  $1.109$ \\
LCDCS 0541     	   &  $1.25$  &  $2.26$  & $0.01$ &  $20.43$          &  $2.0$           & $1.15$  &  $18.52$  & $43.5$ &  $0.793$ \\ 
MACS J1311-0310    &  $1.91$  &  $1.69 $ & $0.00$ &  $19.60$          &  $1.9$           & $0.85$  &  $19.76$  & $28.5$ &  $0.07$ \\  
\hline		
\end{tabular}
\label{tab:nuk_ser}
\end{table*}

In five cases, the fit of the BCG is of better quality when a Nuker
and a S\'ersic function are superimposed, rather than two S\'ersic
laws. For these we give the corresponding Nuker and S\'ersic
parameters in Table~\ref{tab:nuk_ser}. 

We searched the V\'eron-Cetty \& V\'eron (2010) catalogue of quasars
to see if these five BCGs hosted an AGN that could explain why their
profile was so strongly peaked in the centre, but found no
correspondence.

\subsubsection{Modelling with a core-S\'ersic function}

Graham et al. (2003) criticized the Nuker law because of the
instability of its results, and proposed the core-S\'ersic law.  As
the Nuker law, this model comprises two power laws, one for the
internal and one for the external parts of the galaxy, but with the
external power law corresponding to a S\'ersic law:

\begin{equation}
I(r)=I'\left[1+\left(\frac{R_{b}}{r}\right)^{\alpha}\right]^{\gamma/\alpha}exp\left[-b_{n}\left(\frac{r^{\alpha}+R_{b}^{\alpha}}{R_{e}^{\alpha}}\right)^{1/\alpha n}\right]
\end{equation}
\noindent
where $R_{b}$ is the radius where a break occurs between the inner law
(represented by $\gamma$) and the external S\'ersic law, which has an
index $n$ and an effective radius $R_e$. The $b_{n}$ parameter is
defined as in the single S\'ersic law, so $R_e$ is the effective
radius of the outer component and not of the full model. The $\alpha$
parameter characterizes the sharpness of the break and the $I'$ intensity is defined by:

\begin{equation}
I'=I_{b}\ 2^{-\gamma/\alpha}\ exp\left[b_{n}\left(2^{1/\alpha}\ \frac{R_{b}}{R_{e}}\right)^{1/n}\right].
\end{equation}

With the GALFIT-CORSAIR software (Bonfini 2014) we attempted to fit
several BCGs with a core-S\'ersic model, but the fits were 
poor. The difference between the core-S\'ersic and Nuker+S\'ersic
models resides in the supplementary power law in the latter
model. Obviously, this power law is necessary to obtain a good fit, so
we decided not to consider the results of the core-S\'ersic fits.

\section{Variations of the BCG morphological parameters with redshift}

\begin{figure}[h!]
 \begin{center}
\includegraphics[width=6.3cm, angle=0]{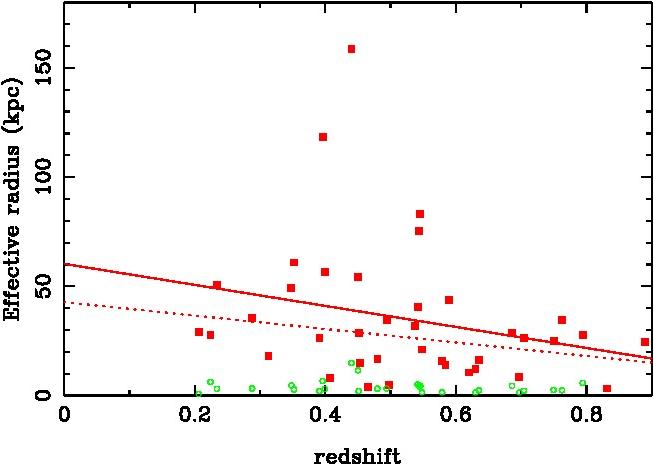}
\caption{Variations of the S\'ersic effective radii of the inner
  (empty green circles) and outer (red squares) components. The error
  bars given by the GALFIT programme are smaller than the symbols. The
  full red line shows the best linear fit when all the red points are
  included, and the dashed line the best fit after eliminating the
  points with an effective radius larger than 70~kpc. }
\label{fig:re_z}
\end{center}
\end{figure}

One of our objectives when analysing this large sample of BCGs was to
check if we could detect a variation of some of the S\'ersic (or
Nuker) parameters with redshift. We indeed find a variation of the
effective radius $R_e$ of the outer component with redshift $z$, as
seen in Fig.~\ref{fig:re_z}. As expected, the effective radius
increases with decreasing redshift, agreeing with the general idea
that BCGs are formed by accreting numerous galaxies in a more or less
continuous way. However, there is a large scatter.  When a linear fit
is made between the effective radius $R_{e}$ and the redshift $z$, for
the outer component we find a slope of $-48.1\pm 29.7$, an intercept
of $60.3\pm 16.3$, a correlation coefficient of $-0.26$ and a
probability that these two quantities are correlated P=89\%.  If the
galaxies with an effective radius larger than 70~kpc are excluded, we
find a slope of $-30.8\pm 14.5$, an intercept of $42.8\pm 8.0$, a
correlation coefficient of $-0.35$ and a probability that these two
quantities are correlated P=96\%. For the inner component, the
probability that these two quantities are correlated is only P=50\%,
so there is no apparent trend.

We should note that in the two clusters with the largest values of
$R_{e,ext}$ extended intracluster light has been detected (MACS0416,
see Montes \& Trujillo 2018, and MACS1206, see DeMaio et al. 2015). In
these two cases, the values of $R_{e,ext}$ are probably overestimated
due to the contribution of the intracluster light (hereafter ICL).

The cosmological dimming factor makes the detection of low surface
brightness features at high redshift difficult. The same source at z=0
and at z=0.8 will be 2.55 magnitudes~arcsec$^{-2}$ fainter at hight
redshift. Consequently, external parts of BCG profiles may be lost at
high redshift.

A way to estimate this effect is to use Fig.~4, where we plot the
effective surface brightness as a function of radius for our
BCGs. Assuming a 2.55 magnitude dimming for a given BCG is equivalent
to reducing the effective radius by 3~kpc to 10~kpc.  If we now look
at Fig.~2, even in the worst case (10 kpc), this is not enough to
explain the decrease of the effective radius between z=0 and z=0.8
only with cosmological dimming effects.

Our results are at least qualitatively in agreement with Ascaso et
al. (2011), who found an increase in the size of BCGs from
intermediate to local redshift.

Lidman et al. (2013) considered a sample of 18 distant clusters with
many spectroscopically confirmed cluster members, and found a major
merger rate of $0.38\pm 0.14$ mergers~Gyr$^{-1}$ at $z\sim 1$.
Assuming that this rate continues to the present day, they find that
it can explain the growth of the stellar mass in BCGs.

\section{Comparison of the orientations of the BCGs and clusters}
\label{sec:orient}

For the 28 clusters for which we analysed both the BCG properties and
the galaxy distribution at very large scale (several Mpc) through
density maps, we now compare the orientations of the BCG and of the
cluster (at the cluster scale or at an even larger scale). For the
remaining ten clusters we could not draw large scale density maps
because we only had small images where the background could not be
estimated sufficiently far from the cluster to estimate the
significance level of the galaxy density.

The method to compute density maps is the following: first, for each
cluster the galaxies located on the cluster red sequence were selected
(the telescope, camera and filter set used for this purpose are
  indicated in the last two columns of Table~\ref{tab:sample}).
We then computedgalaxy density maps for these 
galaxies, based on an adaptive kernel technique with a generalized
Epanechnikov kernel as suggested by Silverman (1986).  Our method is
based on an earlier version developed by Timothy Beers (ADAPT2) and
further improved by Biviano et al. (1996).  The statistical
significance is established by bootstrap resampling of the data. A
density map is computed for each new realisation of the distribution,
with a pixel size of $0.001$~deg (3.6~arcsec).  For each pixel of the
map, the final value is taken as the mean over all realisations. A
mean bootstrapped map of the distribution is thus obtained. The number
of bootstraps used here is 100. More details can be found in the paper
by Durret et al. (2016) from which some density maps are drawn, while
the remaining density maps were computed more recently, based on data
taken from CLASH.

\begin{table}[h]
  \caption{Comparison of the major axis position angle of 28 BCGs (PA$_{\rm BCG}$) and large scale 
    structures, measured anticlockwise from north. 
    For three clusters, the PA of the cluster itself (PA$_{\rm cluster}$) is different from 
    that of the LSS (PA$_{\rm LSS}$), 
    and in these cases both PAs are given. The density maps of the clusters marked with 
    an asterisk have already been published in Durret et al. (2016). The : sign indicates 
    that the PA is not well determined (due to an ellipticity close to 0), and in four
    cases the PA is not given at all when the contours are too close to circular.
  }
\begin{tabular}{llll}
\hline		
\hline		
Cluster      & PA$_{\rm BCG}$ & PA$_{\rm LSS}$ & PA$_{\rm cluster}$\\
\hline		
Cl0016+16$^*$ &~~56 &~~35 & \\
A209          & 146 & 131 & \\ 
Cl0152$^*$    & 134 & 160 & \\
MACS0329      & 158 & 144 & \\ 
MACS0416      &~~40 &~~52 & \\ 
MACS0429      & 167 & 125:& \\ 
MACS0454$^*$  & 113 & 150 & \\
MACS0647$^*$  &~~46 &~~90 & \\
MACS0717$^*$  &~~63 & 122& \\ 
MACS0744$^*$  &~~22 &~~96 & \\ 
A611          &~~38 &~~63 & \\
A851$^*$      &~~76 &~~$-$& \\
LCDCS0172     &~~~~1 & 100& \\
MACS1115      & 148 & 147 & \\ 
MACS1149      & 131 &~~90:& 140 \\ 
MACS1206$^*$  & 104 & 180 & \\ 
BMW-HRI1226$^*$ &~~95 &~~$-$& \\
MACS1311      & 132: & 174 & \\ 
Zw1332$^*$    &~~59 &~~$-$& \\
LCDCS0829$^*$ &~~30 &~~51 & \\ 
MACS1621$^*$  &~~78 & 125 & \\
OC02$^*$      & 126 &~~90 & \\
MACS1720      & 177 &~~49 & 150 \\ 
A2261         & 174 &~~91 &~~60 \\ 
MACS2129$^*$  &~~81 &~~80 & \\
RX2129        &~~64 &~~78 & \\ 
MS2137        &~~71: & 136 & \\ 
RX2328$^*$    & 113 &~~$-$ & \\
\hline		
\end{tabular}
\label{tab:pa}
\end{table}

\begin{table}[h]
  \caption{Maximum extents of the 3$\sigma$ contours of the density maps for 28 clusters. 
    The density maps of the clusters marked with an asterisk have already 
    been published in Durret et al. (2016).}
\begin{tabular}{lrr}
\hline		
\hline		
Cluster  & major axis & minor axis \\
         & (Mpc)      & (Mpc) \\
\hline		
Cl0016+16$^*$& 7.4 & 3.2 \\
A209         & 3.6 & 1.5 \\
Cl0152$^*$   & 2.5 & 2.1 \\
MACS0329     & 3.8 & 2.0 \\
MACS0416     & 4.1 & 2.0 \\
MACS0429     & 2.0 & 1.6 \\
MACS0454$^*$ & 3.9 & 3.4 \\
MACS0647$^*$ & 6.8 & 2.2 \\
MACS0717$^*$ & 6.0 & 1.8 \\
MACS0744$^*$ & 3.8 & 1.5 \\
A611         & 2.3 & 1.2 \\
A851$^*$     & 5.9 & 5.9 \\
LCDCS0172    & 4.9 & 3.2 \\
MACS1115     & 5.1 & 1.7 \\
MACS1149     & 8.6 & 4.0 \\
MACS1206$^*$ & 5.7 & 2.4 \\
BMW-HRI1226$^*$ & 2.3 & 2.0 \\
MACS1311     & 2.4 & 1.4 \\
Zw1332$^*$   & 5.8 & 5.4 \\
LCDCS0829$^*$& 7.5 & 3.3 \\ 
MACS1621$^*$ & 7.6 & 2.1 \\
OC02$^*$     & 6.0 & 4.6 \\
MACS1720     & 2.9 & 2.2 \\
A2261        & 2.9 & 1.8 \\
MACS2129$^*$ & 3.7 & 1.6 \\
RX2129       & 2.6 & 0.9 \\
MS2137       & 1.9 & 1.0 \\
RX2328$^*$   & 1.3 & 1.2 \\
\hline		
\end{tabular}
\label{tab:ext}
\end{table}

We initially intended to measure the major axis position angles of the
BCGs (PA$_{\rm BCG}$) with SExtractor. However, if the inner and outer
isophotes are not elongated along the same PA, the final PA given by
SExtractor is an average between these values. Since we wanted to
compare the elongations of the outer isophotes of the BCGs to the
elongations at the cluster scale or larger, we decided to use for
BCGs the PA given by the IRAF task ELLIPSE, with which we also
computed the BCG light profiles (see Sect.~\ref{sec:method}). The
corresponding PAs are given in Table~\ref{tab:pa}. The values of
PA$_{\rm BCG}$ have typical uncertainties smaller than $\pm
10$~deg.
In the cases when the BCGs appear very round, their PAs are
ill-defined and noted with the : sign in Table~\ref{tab:pa}.

\begin{figure}[h!]
 \begin{center}
\includegraphics[width=6cm, angle=0]{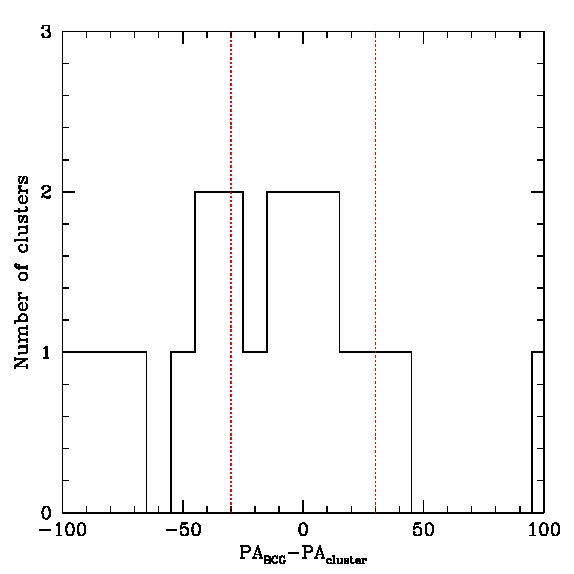}
\caption{Histogram of the differences between the PA of the BCG and
  that of the LSS (or that of the cluster for the three clusters with
  $PA_{cluster}$ given in Table~\ref{tab:pa} for the objects with
  well defined BCG and large-scale structure PAs. The red lines show
  the limits of $\pm 30$~deg within which the PAs are considered to be
  compatible. }
\label{fig:histo_diffPA}
\end{center}
\end{figure}

At very large scale, an indicative ellipse was adjusted by eye to the
3$\sigma$ contours of the density maps (see Durret et al. 2016), and
the major axis position angles of these ellipses (PA$_{\rm LSS}$) are
also given in Table~\ref{tab:pa}.  Since it is necessary to extract
the mean background value in the density map (far from the cluster
region) to compute significance levels, such density maps were only
computed for the clusters of the CLASH or DAFT/FADA surveys for which
large field images (obtained with Subaru/SuprimeCam or CFHT/Megacam)
were available.  In view of the shapes sometimes irregular of the
3$\sigma$ contours, we estimate that the errors on PA$_{\rm LSS}$ can
reach about $\pm 20$~deg. The images of the BCG of Cl0016 and its density maps are shown
in Fig.~\ref{fig:cl0016}.

The sizes of the major and minor axes of the ellipses that were fit to
the $3\sigma$ contours of the large scale density maps are given in
Table~\ref{tab:ext}. We can see in particular the very large extent of
several structures, already noted by Durret et al. (2016), and that of MACS1149, reported here for the first time.

In three cases (MACS1149, RX1720, and A2261), the PA of the cluster
itself (${\rm PA_{cluster}}$) does not coincide with the PA of the
elongation at a larger scale (${\rm PA_{\rm LSS}}$). In this case, we also
give ${\rm PA_{cluster}}$ in Table~\ref{tab:pa}.

The histogram of the values of the difference between the PA of the
BCG and that of the large scale structure (in three cases, the PA of
the cluster itself) is shown in Fig.~\ref{fig:histo_diffPA}. For two
of the three clusters for which ${\rm PA_{LSS}}$ is different from
${\rm PA_{cluster}}$ (MACS1149 and MACS1720) we can note that
${\rm PA_{cluster}}$ is much closer to the value of
${\rm PA_{\rm BCG}}$ than ${\rm PA_{\rm LSS}}$.  Therefore, the values
of PA$_{\rm BCG}$ and either ${\rm PA_{\rm LSS}}$ or
${\rm PA_{cluster}}$ agree within less than 30~deg in 12 cases out of
21 (we count 21 objects, since out of the 28 objects, 7 have at least
one of the two PAs that is ill-defined).

We now briefly consider the objects for which the BCG and larger scale
PAs disagree. For MACS0717, the main cluster has several components,
and PA$_{\rm BCG}$ does not seem to differ very much from that of the
main western component.  The BCGs of
MACS1311 and MS2137 appear very round, so their PA$_{\rm BCG}$ are
probably ill-defined. This is also the case for MACS1149, and besides
the large scale structure shows a large curved extension, so
PA$_{\rm cluster}$ is difficult to estimate. For the other clusters,
the PAs appear well defined but obviously disagree.

In conclusion, out of 28 clusters, if we exclude the three BCGs with
ill-defined PA$_{\rm BCG}$ and the clusters with ill-defined or
undefined PA$_{\rm LSS}$ or PA$_{\rm cluster}$ (in view of their round shape),
we find an agreement of the BCG and large-scale structure PAs within
$\pm 30$~deg for 12 clusters out of 21.

\section{Discussion and conclusions}

\subsection{Kormendy relation and BCG morphological parameters}

\begin{figure}
 \begin{center}
\includegraphics[width=6.3cm, angle=-90]{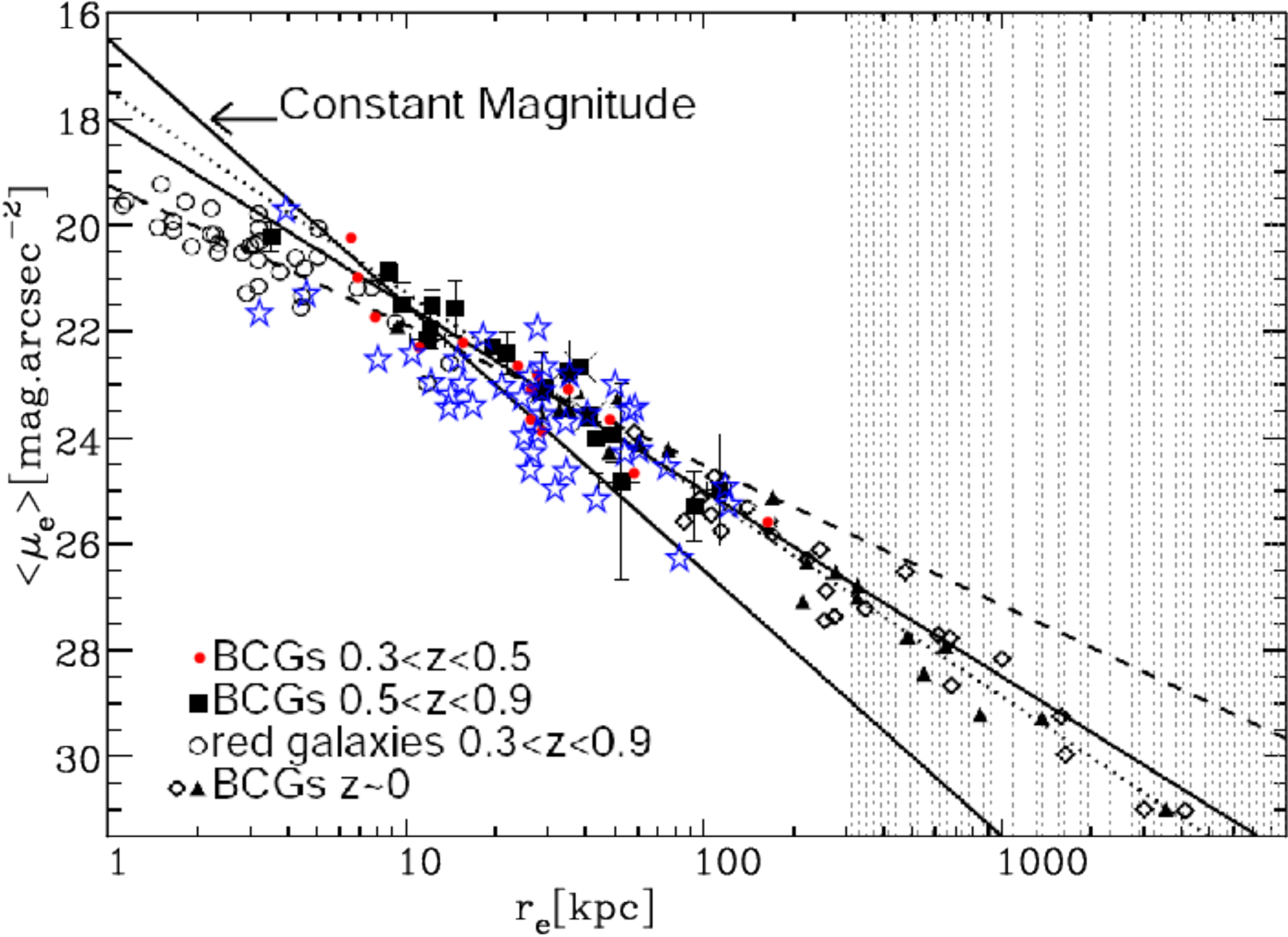}
\caption{Effective surface brightness as a function of effective radius
(Kormendy 1977 relation) drawn by Bai et al. (2014) for BCGs at
various redshifts. We use their Fig. 7 as a reference, where the
shaded region (here drawn with small dotted vertical lines) 
indicates a size that is larger than the BCG images that they used
in their study; the black solid line shows the slope of the constant
magnitude relation (indicated with an arrow); the other black line is
the best fit to their BCG sample; the dotted line is for local
BCGs and the dashed line is the best fit for the lower-mass ETGs. We
superimpose as blue stars our values for the S\'ersic outer component
(all our objects were observed in the F814W filter except one).
}
\label{fig:kormendy_bai}
\end{center}
\end{figure}
\hfill

Fig.~\ref{fig:kormendy_bai} shows the Kormendy (1977) relation drawn
by Bai et al. (2014, see their figure~7) for BCGs at various
redshifts.  We can see that our results for the S\'ersic outer
component fall quite well on this relation, though Bai et al. (2014)
analysed only profiles (instead of 2D structures) and fit them by a
single S\'ersic law.  For our 38 BCGs, the best fit corresponds to a
slope of $2.64\pm 0.35$ and intercept of $19.7\pm 0.5$, with a
correlation coefficient of 0.79.

This can be compared to the relation found by Kormendy (1977):

\begin{equation}
\mu _B = 3.02\ logr_0 + 19.74
\end{equation}
in units of B magnitudes arcsec$^{-2}$. Although the slope is somewhat
different from that of the Kormendy relation,
Fig.~\ref{fig:kormendy_bai} shows that we overall agree.

Bai et al. (2014) found that the masses of the BCGs appear to have
grown by at least a factor of 1.5 from $z=0.5$ to $z=0$, in contrast
to previous findings of no evolution, and argued that such an
evolution validates the expectation from the $\Lambda$CDM model. Since
our results are consistent with theirs, we believe that this
strengthens their conclusions. These results also agree with Burke \&
Collins (2013) who counted galaxies around the BCGs of 14 clusters in
the redshift range $0.8<z<1.4$ and found that the BCG stellar mass
could have increased by as much as a factor of 1.8 between $z=1$ and
the present epoch.

We have seen in Fig.~\ref{fig:re_z} that the effective radius of BCGs
increases with decreasing redshift. This agrees with models of BCG
formation and evolution (e.g. Arag\'on-Salamanca et al. 1998, De Lucia
\& Blaizot 2007), who found that BCGs assemble quite late (half their
final mass is typically locked up in a single galaxy after z$\sim$0.5).

In a future work, we plan to estimate the masses of our 38 BCGs, to
quantify the growth in mass with decreasing redshift for our
sample. It would also be interesting to see if the offset of the BCG
position relative to the cluster centre is correlated to the degree of
concentration of cluster X-ray morphology, and to see if the brighter
BCGs are preferentially found in morphologically disturbed clusters,
as done by Hashimoto et al. (2014), based on ground-based data
obtained with Subaru.

\subsection{Preferential orientations}

In Sect.~\ref{sec:orient}, we have compared the PAs of the structures
at different scales: PA${\rm _{BCG}}$, PA${\rm _{cluster}}$, and
PA${\rm _{LSS}}$. We found an agreement of PA$_{\rm BCG}$ and
PA$_{\rm LSS}$ (or in three cases PA$_{\rm cluster}$) within $\pm 30$~deg
for 12 clusters out of 21 (excluding BCGs or clusters where the PAs
are ill-defined or undefined). In view of recent results by West et
al. (2017), we expected that the PAs would agree for a larger number
of clusters. Based on Hubble Space Telescope observations of 65
distant galaxy clusters, these authors found that giant elliptical
galaxies in the centres of rich clusters often have major axes sharing
the same orientation as the surrounding matter distribution on larger
scales. They argued that BCGs are the product of a special formation
history, influenced by the development of the cosmic web over billions of
years.  At lower mass, Paz et al. (2011) found an alignment of galaxy
groups with the surrounding large scale structure, with a strong
alignment signal between the projected major axis of group shapes and
the surrounding galaxy distribution up to scales of 30~Mpc~h$^{-1}$,
this observed anisotropy signal becoming larger as the galaxy group
mass increases.

Looking at the figures of Appendix B (right column), we can see
that in 6 cases out of 17 (MACS0329, MACS0416, MACS1149, LCDCS0829,
RX2129, and MS2137) the main cluster is elongated towards the other
nearby structures. Plionis \& Basilakos (2002) stated that clusters
were aligned towards their nearest neighbour, specially within
superclusters, suggesting anisotropic merging. This is probably the
case in the six above-mentioned clusters, which also have properties
comparable to those of the three intermediate redshift clusters
studied by Fo\"ex et al. (2017). These authors found that the optical
morphology of the clusters correlates with the orientation of their
BCG, and with the position of the main axes of accretion.

\subsection{Relation between the BCG and the ICL}

\begin{figure}
 \begin{center}
\includegraphics[width=6.3cm, angle=0]{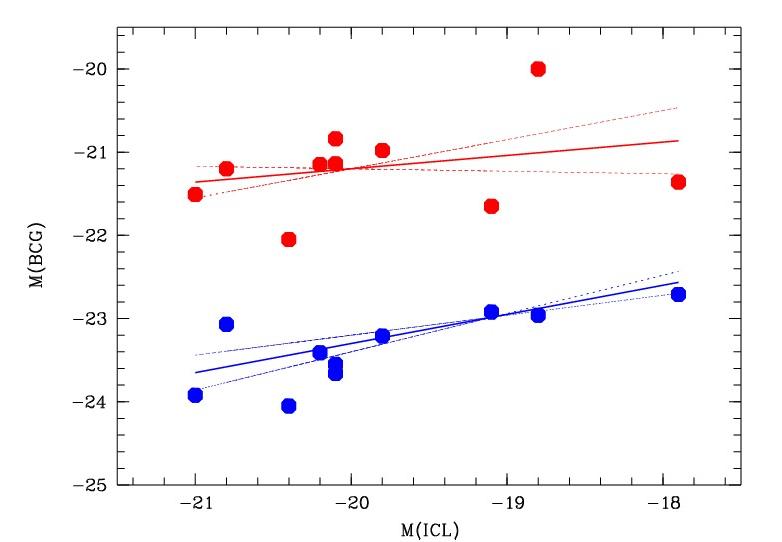}
\caption{Absolute magnitude of the outer (blue points) and inner (red
  points) components of the BCG as a function of the absolute
  magnitude of the intracluster light for the ten clusters in common
  with Guennou et al. (2012). The corresponding regressions are shown
  as full lines and the regressions within $\pm 1\sigma$ of the best
  fit parameters as dotted lines (see text). }
\label{fig:magbcg_magicl}
\end{center}
\end{figure}

Comparing our results (Table~2) with the work of Guennou et
al. (2012), we first note that all ten clusters of Guennou et
al. (2012) have detected intracluster light (ICL) and that they are
all better fit with two S\'ersic functions. This could suggest that
physical processes having created the ICL are also at the origin of an
extended halo of the BCGs.  In order to investigate further this
possible relation, we compared for these ten clusters the total
absolute magnitude of the external S\'ersic component of the BCG
$\rm M_{BCG,ext}$ (see Table~2) with the total absolute magnitude of the
cluster ICL $\rm M_{ICL}$ from Guennou et
al. (2012). Fig.~\ref{fig:magbcg_magicl} shows the correlation between these
two components, the best fit being:

$$\rm M_{BCG,ext} = (0.35\pm 0.11) M_{ICL} - (16.3\pm 2.1).$$

As a comparison, we perform the same exercise with the total absolute magnitude of the internal S\'ersic component of the BCG, and as expected we have
no significant relation:

$$\rm M_{BCG,int} = (0.16\pm 0.19) M_{ICL} - (18.0\pm 3.8).$$

To explain this relation, one may argue that the ICL detected by
Guennou et al. (2012) is simply a part of the external BCG halo for
each cluster.  However, the external BCG halos are much brighter than
the detected ICL, so we are not considering the same light sources.
Moreover, ICL sources are extending up to 60~kpc from the cluster
centers (see Fig. 5 of Guennou et al. 2012), while the effective radii
of Table~2 are in most cases lower than 30~kpc. We are therefore
not sampling the same cluster areas. We therefore propose that the
physical phenomena at the origin of the ICL are related to the
formation of the BCG halos, both qualitatively and quantitatively.

However, based on CLASH data, Burke et al. (2015) claim that the
ICL and BCG were not built up by the same mechanism. They found that
minor mergers (mergers with objects with masses half of the BCG
mass) are the dominant process for stellar mass assembly at low
redshifts, the majority of the stellar mass from interactions
contributing to the ICL, rather than building up the BCG. Therefore,
their point of view is that different processes build up the ICL and
BCGs. We must however note that they do not extract the ICL
contribution in the same way as Guennou et al. (2012), so the
results of these two papers may not be not directly comparable.

\subsection{Conclusions}

Our study is limited here to redshifts $z<0.9$. It would be
interesting to study BCGs at larger redshifts to see if their growth
can be traced at higher redshifts. Concerning the alignments of the
major axes of BCGs with the elongations of larger scale structures,
the study of a statistically significant sample of clusters in the
present redshift range is now timely to reach conclusive results, and
observations at $z>1$ would be of interest to check if the properties
discussed here are also observed in the earlier universe. The
  implications of such a study for testing the cluster formation and
  evolution paradigm clearly requires larger samples and a proper
  comparison with cosmological simulations, and this will be the
  object of future work.

\begin{acknowledgements}

  We are very grateful to Patrick Hudelot for his help in reducing
  some CFHT/MegaCam images and to Tabatha Sauvaget and Nicolas
  Martinet for discussions on GALFIT. We thank the referee for
  interesting comments.  F.D. acknowledges long-term support from
  CNES. I.M. acknowledges support from the Spanish Ministry of Economy
  and Competitiveness through grants AYA2013-42227-P and
  AYA2016-76682-C3-1-P.

  The scientific results reported in this article are based on
  publicly available HST data acquired with ACS through the CLASH and
  COSMOS surveys, and on Subaru Suprime-Cam archive data collected at
  the Subaru Telescope, which is operated by the National Astronomical
  Observatory of Japan. Also based on observations made with the FORS2
  multi-object spectrograph mounted on the Antu VLT telescope at
  ESO-Paranal Observatory (programmes 085.A-0016, 191.A-0268; PI: C.
  Adami). Also based on observations obtained at the Gemini
  Observatory, which is operated by the Association of Universities
  for Research in Astronomy, Inc., under a cooperative agreement with
  the NSF on behalf of the Gemini partnership: the National Science
  Foundation (United States), the Science and Technology Facilities
  Council (United Kingdom), the National Research Council (Canada),
  CONICYT (Chile), the Australian Research Council (Australia),
  Minist\'erio da Ci\^encia, Tecnologia e Inova\c cao (Brazil), and
  Ministerio de Ciencia, Tecnolog\'\i a e Inovaci\'on Productiva
  (Argentina). Also based on observations made with the Italian
  Telescopio Nazionale Galileo (TNG) operated on the island of La
  Palma by the Fundaci\'on Galileo Galilei of the INAF (Istituto
  Nazionale di Astrofisica) at the Spanish Observatorio del Roque de
  los Muchachos of the Instituto de Astrof\'\i sica de Canarias. Also
  based on service observations made with the WHT operated on the
  island of La Palma by the Isaac Newton Group in the Spanish
  Observatorio del Roque de los Muchachos of the Instituto de
  Astrof\'\i sica de Canarias. Also based on observations collected at
  the German- Spanish Astronomical Center, Calar Alto, jointly
  operated by the Max-Planck- Institut fur Astronomie Heidelberg and
  the Instituto de Astrof\'\i sica de Andaluc\'\i a (CSIC). Based on
  observations obtained with MegaPrime/MegaCam, a joint project of
  CFHT and CEA/IRFU, at the Canada-France-Hawaii Telescope (CFHT)
  which is operated by the National Research Council (NRC) of Canada,
  the Institut National des Sciences de l'Univers of the Centre
  National de la Recherche Scientifique (CNRS) of France, and the
  University of Hawaii.  This work is partly based on data products
  produced at Terapix available at the Canadian Astronomy Data Centre
  as part of the Canada-France-Hawaii Telescope Legacy Survey, a
  collaborative project of NRC and CNRS. Also based on observations
  obtained at the WIYN telescope (KNPO). The WIYN Observatory is a
  joint facility of the University of Wisconsin-Madison, Indiana
  University, Yale University, and the National Optical Astronomy
  Observatory.  Kitt Peak National Observatory, National Optical
  Astronomy Observatory, is operated by the Association of
  Universities for Research in Astronomy (AURA) under cooperative
  agreement with the National Science Foundation.  Also based on
  observations obtained at the MDM observatory (2.4 m telescope).  MDM
  consortium partners are Columbia University Department of Astronomy
  and Astrophysics, Dartmouth College Department of Physics and
  Astronomy, University of Michigan Astronomy Department, The Ohio
  State University Astronomy Department, Ohio University Dept. of
  Physics and Astronomy.  Also based on observations obtained at the
  Southern Astrophysical Research (SOAR) Telescope, which is a joint
  project of the Minist\'erio da Ci\^encia, Tecnologia, e Inova\c cao
  (MCTI) da Rep\'ublica Federativa do Brasil, the US National Optical
  Astronomy Observatory (NOAO), the University of North Carolina at
  Chapel Hill (UNC), and Michigan State University (MSU). Also based
  on observations obtained at the Cerro Tololo Inter-American
  Observatory, National Optical Astronomy Observatory, which are
  operated by the Association of Universities for Research in
  Astronomy, under contract with the National Science
  Foundation. Finally, this research has made use of the VizieR
  catalogue access tool, CDS, Strasbourg, France and of the NASA/IPAC
  Extragalactic Database (NED), which is operated by the Jet
  Propulsion Laboratory, California Institute of Technology, under
  contract with the National Aeronautics and Space Administration.

\end{acknowledgements}

\clearpage

\appendix

\section{Residual and sharp-divided maps}

\begin{figure*}
\begin{center}
\mbox{\psfig{figure=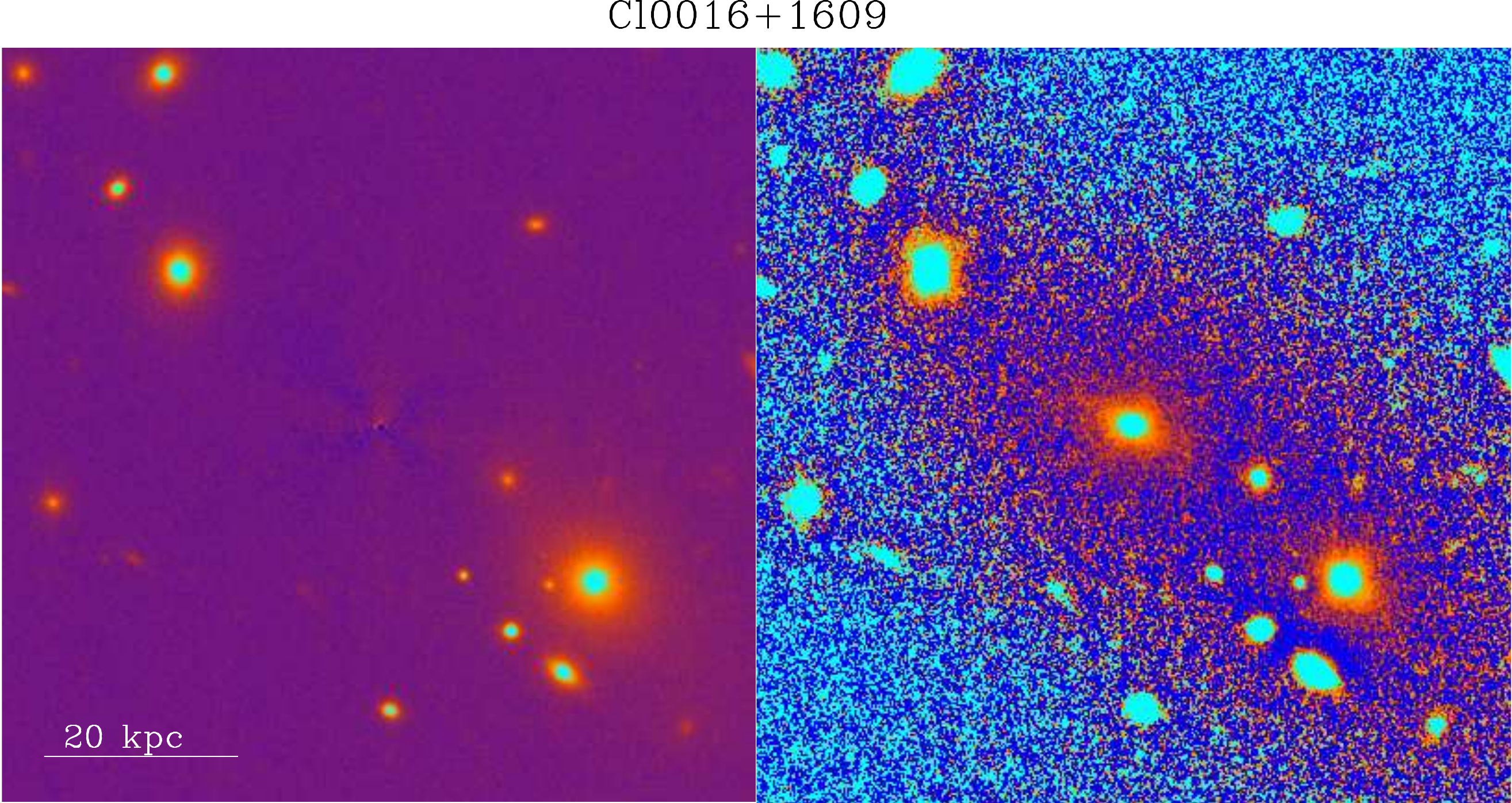,width=8.5cm,angle=0}}
\caption{Left: residuals after fitting Cl0016 with two S\'ersic
  profiles. Right: sharp divided image of the BCG of Cl0016. The two
  images have identical scales. North is top and East is left. }
\label{fig:cl0016rsd}
\end{center}
\end{figure*}

In Fig.~\ref{fig:cl0016rsd}, we show for the BCG of Cl0016
analysed with GALFIT the map of the residuals obtained after
subtracting to the BCG its best fit (by the sum of two S\'ersic models),
and the corresponding sharp divided image, in the F814W band. 
The sharp divided image was mainly used to identify
correctly the objects that needed to be masked in order to obtain the
best possible fit of the BCGs. The maps of the residuals obtained by subtracting 
to the BCGs its best fits (by a single S\'ersic model, by the sum of two S\'ersic models,
or by the sum of a S\'ersic and a Nuker models) for all the clusters of our sample and the corresponding sharp divided images
are available on this link : \url{ftp://ftp.iap.fr/pub/from_users/durret/BCGs/Durret_BCGs.pdf}

Cl0016: the residuals are very faint, showing that the fit is
good. The sharp divided image seems to show a diffuse halo around the
BCG.

A209: the residuals show matter in the very centre of the BCG as well
as in its outskirts.

Cl0152: here also there is matter left in the central part of the
galaxy.

MACS0329: a relatively bright feature extends up to about 20~kpc
north-west of the BCG centre.

MACS0416: some large scale diffuse emission is seen around the BCG.

MACS 0429: besides possible diffuse light at large scale, there are many 
features in the BCG area.

MACS0454: the fit is very good except in the very central regions of
the BCG. 

MACS0647: an excess is visible near the BCG centre.

MACS 0717: an excess is visible near the BCG centre.

MACS0744, A611, A851, LCDCS0110, LCDCS0130: same as Cl0016.

LCDCS0172: the fit to the BCG is not perfect, probably due to the
presence of a small galaxy a few kpc north of the BCG.

LCDCS0173, Cl1103, MACS1115, LCDCS0340: same as Cl0016.

MACS1149: an elongated emission region crosses the BCG in the
north-west to south-east direction.

MACS1206, LCDCS0504, BMW1226, LCDCS0531: same as Cl0016.

LCDCS0541: the fit is not perfect.

MACS1311: same as Cl0016.

Zw1332: the fit is good, and the sharp divided image reveals faint
features south-west and north-east of the BCG centre.

LCDCS0829 (RX1347), LCDCS0853: the fit is not perfect, and large diffuse emission
is visible.

MACS1621: the fit is very good.

OC02: the fit is not perfect, and some diffuse emission is visible.

MACS1720: same as Cl0016.

A2261: same as Cl0016 though the fit is not perfect.

MACS2129: the fit is not perfect.

RX2129, MS2137, RX2248: same as Cl0016.

RX2328: an elliptical residual is clearly seen in the central zones of
the BCG

\clearpage

\section{Comparison of the orientations of the BCGs and clusters}

\begin{figure*}
 \begin{center}
\mbox{\kern-2.6cm\psfig{figure=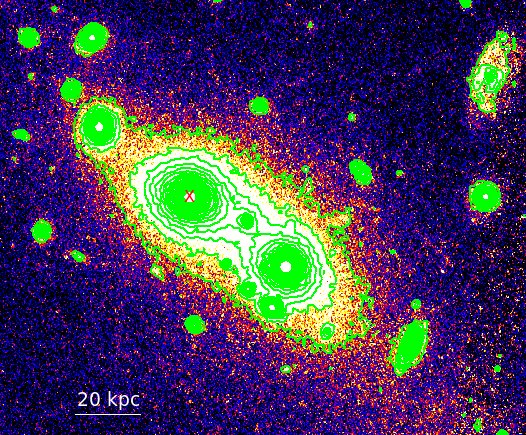,width=7.0truecm,angle=0}\kern0.2cm%
\psfig{figure=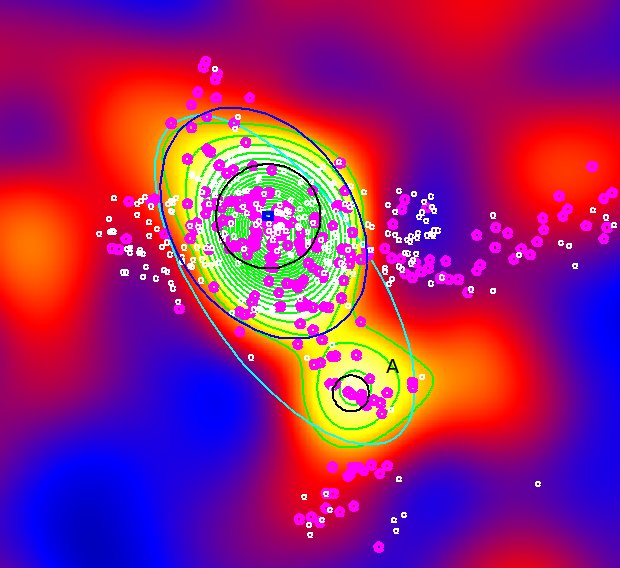,width=6.3truecm,angle=0}}
\caption{Cl0016+1609. Left: image of the BCG. The BCG is indicated
  with a red cross, the small galaxy to the north-east belongs to the
  cluster, but not the large galaxy located south-west of the
  BCG. Right: large scale density map computed from the galaxies
  located on the cluster red sequence.  The black circle is centred on
  the position of the cluster given in Table~1 and has a 1~Mpc radius,
  as in all following figures (in some cases, the circle is drawn in
  another colour to be more visible). In both figures, the contour
  levels start at $3\sigma$, they increase by $2\sigma$ or $3\sigma$
  in the left figures, depending on the brightness of the BCG, and by
  $1\sigma$ in the density maps.  In the right figure, the ellipse
  (here in cyan, but drawn with different colours for other clusters
  to be clearly visible) indicates the maximum extent of the $3\sigma$
  contours. The small green or white points show the positions of the
  galaxies with a measured spectroscopic redshift. The magenta points
  show the galaxies with redshifts in the approximate cluster redshift
  range (this range is indicated in the figure caption of each
  cluster, it is $0.53<z<0.57$ for Cl0016+1609), and the small blue
  rectangle at the centre shows the size of the left figure (in this
  and following figures). North is up and east is left in all figures.
}
\label{fig:cl0016}
\end{center}
\end{figure*}

In Fig.~\ref{fig:cl0016} we show for the cluster Cl0016
the image of the BCG and of the large scale structure around the
cluster. As mentioned in Appendix A, these figures for all the 
clusters are available on this website : \url{ftp://ftp.iap.fr/pub/from_users/durret/BCGs/Durret_BCGs.pdf}.
Some of the large scale density maps have already been
published in Durret et al. (2016) but we show them again for easy
comparison with the BCG. In some cases, we zoomed them to show only
the cluster and its immediate surroundings. Some regions labeled on
the density maps (A, B...) are the same as in Durret et al. (2016).

\end{document}